\documentclass[10pt,twocolumn,letterpaper,superscriptaddress]{revtex4}

\usepackage{multibbl}
\usepackage[pdftex]{graphicx}
\usepackage{amsmath,SIunits}
\usepackage[latin1]{inputenc}
\usepackage{latexsym,stmaryrd}
\usepackage[sort&compress]{natbib}
\usepackage[pdftex]{graphicx}
\usepackage{color}
\bibpunct{[}{]}{,}{n}{}{,}

\makeatletter
\newcommand*{\citenst}[2][]{%
  \begingroup
  \let\NAT@mbox=\mbox
  \let\@cite\NAT@citenum
  \let\NAT@space\NAT@spacechar
  \let\NAT@super@kern\relax
  \renewcommand\NAT@open{[}%
  \renewcommand\NAT@close{]}%
  \citep{#2}%
  \endgroup
}
\makeatother

\begin{document}

\title{Coherent generation and detection of acoustic phonons in topological nanocavities}

\author{G. Arregui}
\altaffiliation{Contributed equally to this work}
\affiliation{Catalan Institute of Nanoscience and Nanotechnology (ICN2), CSIC and BIST, Campus UAB, Bellaterra, 08193 Barcelona, Spain}
\affiliation{Dept. de F\'{i}sica, Universitat Autonoma de Barcelona, 08193 Bellaterra, Spain}
\author{O. Ort\'{i}z}
\altaffiliation{Contributed equally to this work}
\affiliation{Centre de Nanosciences et de Nanotechnologies (C2N), CNRS, Univ. Paris-Sud, Universit\'e Paris-Saclay, 91120 Palaiseau, France}
\author{M. Esmann}
\affiliation{Centre de Nanosciences et de Nanotechnologies (C2N), CNRS, Univ. Paris-Sud, Universit\'e Paris-Saclay, 91120 Palaiseau, France}
\author{C. M. Sotomayor-Torres}
\affiliation{Catalan Institute of Nanoscience and Nanotechnology (ICN2), CSIC and BIST, Campus UAB, Bellaterra, 08193 Barcelona, Spain}
\affiliation{ICREA - Instituci\'o Catalana de Recerca i Estudis Avan\c{c}ats, 08010 Barcelona, Spain}
\author{C. Gomez-Carbonell}
\affiliation{Centre de Nanosciences et de Nanotechnologies (C2N), CNRS, Univ. Paris-Sud, Universit\'e Paris-Saclay, 91120 Palaiseau, France}
\author{O. Mauguin}
\affiliation{Centre de Nanosciences et de Nanotechnologies (C2N), CNRS, Univ. Paris-Sud, Universit\'e Paris-Saclay, 91120 Palaiseau, France}
\author{B. Perrin}
\affiliation{Institut de Nanosciences de Paris, Sorbonne Universit\'e, CNRS UMR 7588, 75005 Paris, France}
\author{A. Lema\^itre}
\affiliation{Centre de Nanosciences et de Nanotechnologies (C2N), CNRS, Univ. Paris-Sud, Universit\'e Paris-Saclay, 91120 Palaiseau, France}
\author{P. D. Garc\'{i}a}
\email{david.garcia@icn2.cat}
\affiliation{Catalan Institute of Nanoscience and Nanotechnology (ICN2), CSIC and BIST, Campus UAB, Bellaterra, 08193 Barcelona, Spain}
\author{N. D. Lanzillotti-Kimura}
\email{daniel.kimura@c2n.upsaclay.fr}
\affiliation{Centre de Nanosciences et de Nanotechnologies (C2N), CNRS, Univ. Paris-Sud, Universit\'e Paris-Saclay, 91120 Palaiseau, France}

\date{\today}

\small

\maketitle

\textbf{Inspired by concepts developed for fermionic systems in the framework of condensed matter physics, topology and topological states are recently being explored also in bosonic systems~\cite{PRXMarquardt}. The possibility of engineering systems with unidirectional wave propagation and protected against disorder is at the heart of this growing interest~\cite{Snowflake}. Topogical acoustic effects have been observed in a variety of systems~\cite{Xiao,HeTopologicalInsulator,NashPNAS,YangRevTopAcoustics,LuNatPhys,YuNatComm}, most of them based on kHz-MHz sound waves, with typical wavelength of the order of the centimeter. Recently, some of these concepts have been successfully transferred to acoustic phonons in nanoscaled multilayered systems~\cite{esmann,esmann2}. The reported demonstration of confined topological phononic modes was based on Raman scattering spectroscopy~\cite{esmann}, yet the resolution did not suffice to determine lifetimes and to identify other acoustic modes in the system. Here, we use time-resolved pump-probe measurements~\cite{thomsen,thomsen2,ultrasonicsruellogusev} using an asynchronous optical sampling (ASOPS) technique~\cite{bartelsASOPS} to overcome these resolution limitations. By means of one-dimensional GaAs/AlAs distributed Bragg reflectors (DBRs) as building blocks~\cite{FPcavity,phonmolecules,ultrasonicsKimura,starkladder,MgOSLs}, we engineer high frequency ($\sim$ 200 GHz) topological acoustic interface states~\cite{esmann,XiaoPRX}. We are able to clearly distinguish confined topological states from stationary band edge modes. The detection scheme reflects the symmetry of the modes directly through the selection rules~\cite{pascualwinter1,generationdetectionrules}, evidencing the topological nature of the measured confined state. These experiments enable a new tool in the study of the more complex topology-driven phonon dynamics such as phonon nonlinearities and optomechanical systems with simultaneous confinement of light and sound~\cite{doublemagiccoincidence,AL}.}

\section{Introduction}

The atomic monolayer accuracy of molecular beam epitaxy (MBE) allows nanostructures based on acoustic impedance modulation in the growth direction, resulting in very precise control of sub-THz mechanical motion. Spectral and spatial tailoring of the acoustic excitations supported in these nanostructures mostly relies on exploiting the energy band structure of periodic one-dimensional phononic crystals. The artificial periodicity leads to frequency band gaps within which elastic waves cannot propagate~\cite{tamura}. Distributed Bragg reflectors (DBRs), the finite version of the infinite crystal, act as mirrors for frequencies inside the band gap of the underlying crystal and constitute the building block for the standard Fabry-Perot type acoustic cavity~\cite{FPcavity}. This confinement approach is based on a spacer layer, bounded by two equal DBRs, leading to a confined mode which exponentially decays along the mirrors, in an analogous manner to the electronic wave function in a quantum well. Recently, another type of cavity based on an adiabatic confinement potential rather than a spacer has been proposed~\cite{adiabaticcavity}. These two strategies solely use the frequency band structure of their underlying periodic counterpart as a confinement strategy, the spatial distribution of the Bloch modes having no particular role in the control of the density of states.\\

A recent work~\cite{esmann} has evidenced a completely novel approach to confine acoustic vibrations in multilayered structures, based on topological band inversion in GaAs/AlAs superlattices. The idea is based on the fact that two concatenated semi-infinite superlattices exhibiting a common band gap region and having inverted symmetries of the Bloch modes at the minigap edges will give rise to an interface state in the band gap region~\cite{XiaoPRX}. The finite-size version of such structure (two concatenated DBRs) will inherit the presence of this topological interface state. The experimental evidence of such states given so far is based on Raman scattering measurements, with limited spectral resolution. In this paper we present time-resolved pump-probe differential reflectivity experiments. Optical generation and detection of acoustic vibrations with ultrafast pulsed lasers constitutes a widely used approach to access complex wave dynamics and the modal structure of acoustic nanoresonators~\cite{hudert,bartelsGaAs,doublemagiccoincidence,micropillars,danimetasurfaces,danimetasurfaces2,nanoletterswright,nanolettersbragas,nanolettersorrit}. In such experiments, tailoring of the generated and detected spectrum can be achieved by design of both sample structure and experimental conditions~\cite{mfpascualwinter,trigo2}. Here, we unveil the detailed structure of  nanoacoustic modes, allowing a clear assignment of the peaks to topological and other stationary modes.\\

\begin{figure*} \includegraphics[width=\textwidth]{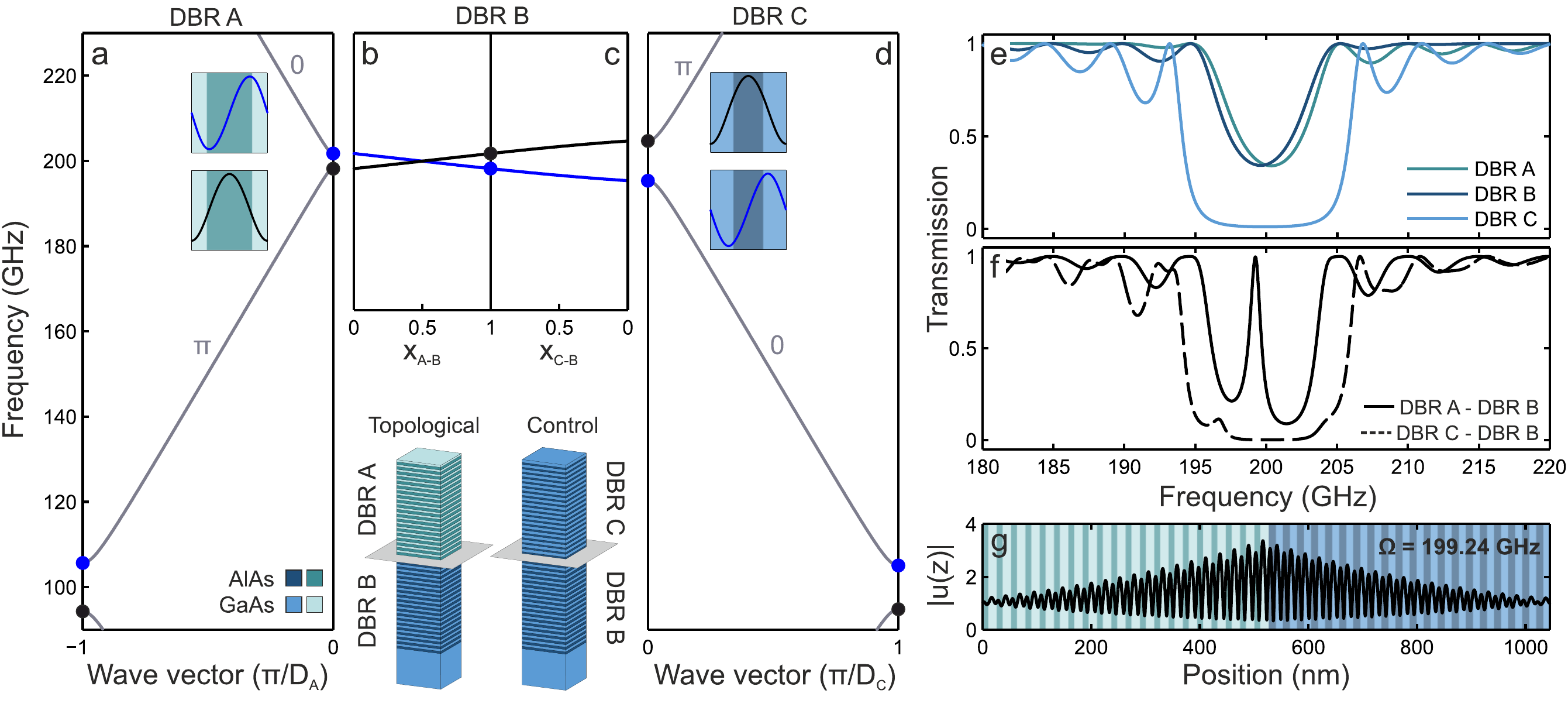}
    \caption{ \label{1} \textbf{Topological interface state through band inversion in GaAs/AlAs superlattices.} (a) Band structure corresponding to DBR A. Zak phases values of different bands depicted on top of them. Minigap edges at the center and border of the Brillouin zone are marked with blue and black dots. Black (blue) dots represent edge modes with a symmetric (anti-symmetric) displacement profile as shown in the insets. (b) and (c) present plots of the edges position as a function of parameters $x_{A-B}$ and $x_{C-B}$, respectively. Edges corresponding to DBR B are represented as dots at $x_{A-B}=x_{C-B}=1$ using the same symmetric/antisymmetric color code for both lines and dots. (d) Band structure corresponding to DBR C using the same representation used in (a). (e) Reflectivity spectra for DBR A, B and C. (f) Reflectivity spectra for topological and control sample. (g) Displacement profile of the topological interface state.}
\end{figure*}

\section{Topological nanophononic resonators}

We  studied two different nanoacoustic multilayer structures using three different types of DBRs (labelled A, B and C) as building blocks. Each DBR contained 20 GaAs/AlAs layer pairs with 10.76nm/15.57nm, 13.15nm/12.74nm and 15.54nm/9.91nm nominal thickness, respectively. Both samples were grown on a 200 $\mu$m thick GaAs substrate  starting with a DBR of type B followed by DBR of type A (we label this the topological sample) or type C (control sample), respectively. While the topological sample is designed to confine a topological acoustic mode at the interface between the two DBRs based on band inversion, the control sample merely supports extended propagating and stationary band edge modes. This difference in the topological properties of the two samples can be understood from an analysis of the DBR bandstructures and the associated mode symmetries  \cite{Xiao,esmann} as shown in Fig. 1(a-d). The band structures of the DBRs A and C are displayed in Fig. 1(a) and 1(d), respectively, with common acoustic minigaps centered around 100GHz and 200GHz. For each acoustic band, the spatial Bloch mode symmetries at the edge and the center of the first Brillouin zone are indicated with black (symmetric) and blue (anti-symmetric) dots. It has been shown, that a Zak phase (equivalent of a Berry's phase in one-dimensional periodic systems) can be associated to each band. It acquires a value of $\pi$ if the mode symmetries at the edge and the center of the Brillouin zone are opposite (second band of DBR A), otherwise the topological phase is $0$ (second band of DBR C). As a consequence, the mode symmetries around the second acoustic minigap have an inverted energetic order (see insets to panels a and d) allowing the formation of an interface state upon concatenation of the two DBRs \cite{esmann}. In Fig. 1(b) and 1(c) we illustrate that upon a topological phase transition the energetic order of the band edge modes changes and we compare the topological phase of DBR B to that of A and C. To that end, we define two continuous parameters $x_{A-B}$ ($x_{C-B}$) that deform DBR A (DBR C) into DBR B. That is, we define the thickness of a DBR's GaAs layers as $d=d_{Y}\cdot(1-x_{Y-B})+d_{B}\cdot x_{Y-B}$ and of the AlAs layers as $e=e_{Y}\cdot(1-x_{Y-B})+e_{ B}\cdot x_{Y-B}$ with $x_{Y-B}\in[0,1]$, $d_{Y} $ and $e_{Y}$ being the nominal sizes of the layers of DBR Y, and $Y$ being A or C. In Fig. 1(b) and 1(c) we plot the evolution of the band edges bounding the second acoustic minigap as a function of $x_{A-B}$ (b) and $x_{C-B}$ (c). As depicted in Fig. 1(c), continuously transforming DBR A into B merely implies a change in the size of the common minigap, but the energetic order of mode symmetries persists. In contrast, Fig. 1(b) shows that a continuous transition between DBR B and C necessarily implies a band crossing, i.e. the minigap closes and re-opens, and an associated exchange of the band edge symmetries. DBRs B and C are hence in the same topological phase, whereas DBRs B and A cannot be continuously transformed into one another, that is, they are in different topological phases.

Using a transfer matrix simulation, we furthermore compute acoustic transmission spectra of the individual DBRs (Fig. 1(e)) and the two concatenated structures (Fig. 1(f)). For each individual DBR we find a broad dip in transmission centered around 200 GHz, which directly reflects the position and size of the common acoustic minigap. For the control sample (dashed line, panel f) only a broad stop band as for the individual DBRs is found. For the topological sample (solid line), however, a clear peak appears at $199.24$ GHz, indicating the presence of a confined mode. The corresponding spatial acoustic displacement profile $\lvert u(z) \rvert$ is shown in Fig. 1(g) superimposed with the layer structure of the topological sample. We observe that the mode is indeed centered at the interface between the two DBRs and decays exponentially to both extremes of the structure. For the control sample, no occurrence of a topological interface mode is expected.

Both samples were grown by MBE on a [001]-oriented GaAs substrate and pre-characterized by means of high resolution x-ray diffraction (HRXRD). As an important tool for structural characterization, HRXRD provides valuable information on the periodicity, layer sizes and overall quality of both samples. A $\theta-2\theta$ HRXRD scan using Cu K-$\alpha$ 1 radiation was performed, diffractograms were measured and further analysis of their peaks provided information about the different parts of the structure. For the topological sample, the results showed that DBR A is formed by GaAs/AlAs layers of 11.06/15.48 nm, whereas DBR B of 13.52/12.67 nm. For the control sample DBR C presents layers of 15.98/9.96 nm of GaAs/AlAs while DBR B of 13.52/12.8 nm. For the topological sample, these thickness values represent a reduction of the AlAs layers by 0.6$\%$ of their nominal values, while for the GaAs layers the change corresponds to an increase of 2.8$\%$. For the control sample both AlAs and GaAs layers present an increase of 0.5$\%$ and 2.8$\%$ respectively. Despite this deviation from the design values, the band structure analysis for both samples and, as consequence, their predicted phonon dynamics remain valid as we show in the following section.

\begin{figure} \includegraphics[width=\columnwidth]{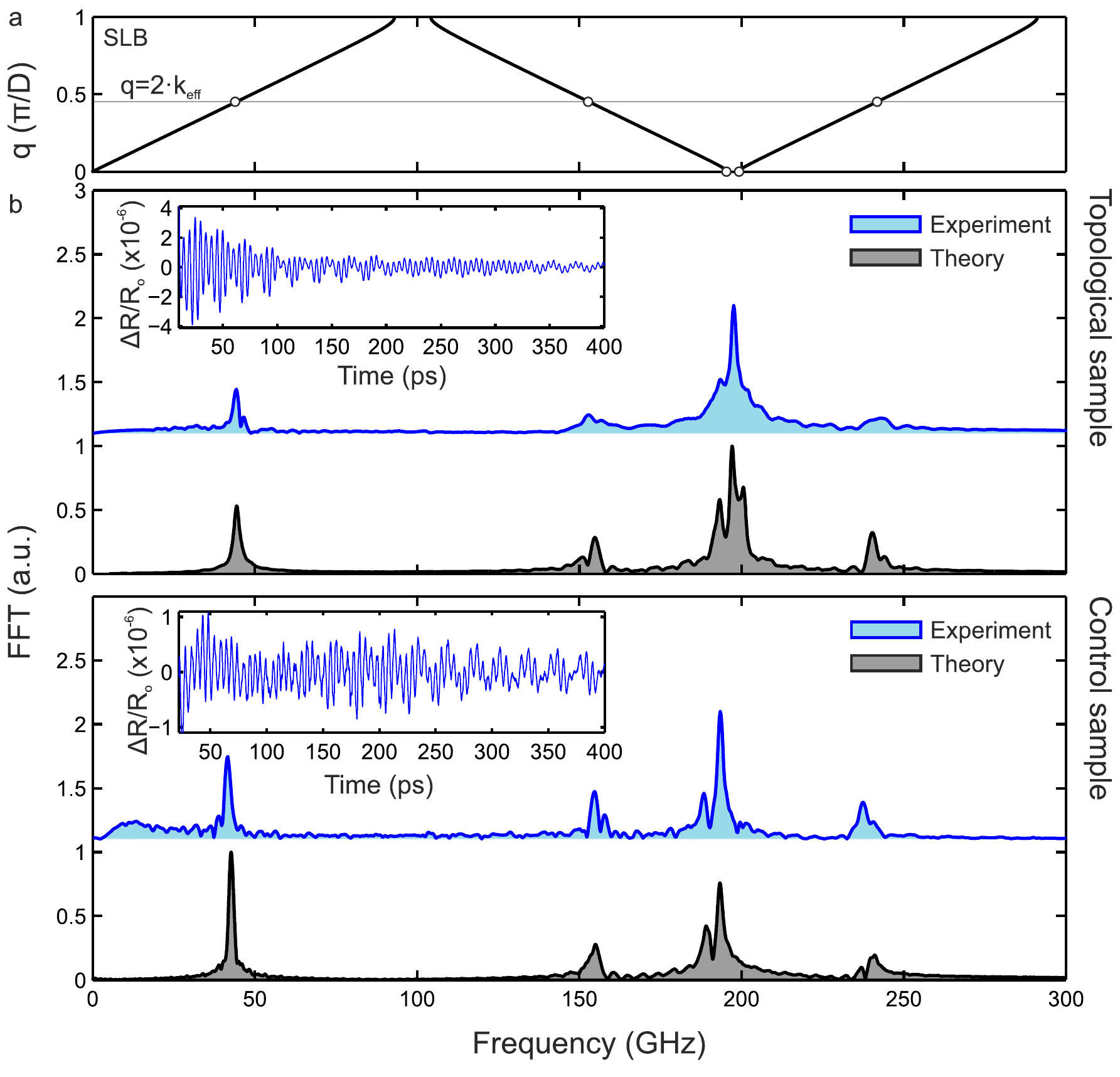}
    \caption{ \label{2} \textbf{Reflection-type pump-probe coherent acoustic phonon experiments.}(a) Band diagram of one of the building superlattices (SL B) with the frequencies related to $q=0$ and $q = 2k_{eff}$ highlighted with circles. (b) Differential reflectivity time-traces (insets) obtained using and asynchronous optical sampling (ASOPS) technique on the topological (top) and control (bottom) samples. The used measurement conditions for pump and probe lasers are 800 nm (40 mW) and 765 nm (4 mW) for the topological sample and 800 nm (40 mW) and 800 nm (4 mW) for the control sample. The as-obtained data was treated by cutting the initial electronic peak and using appropiate Savitzy-Golay filtering to extract low-frequency backgrounds. The Fourier transform of the experimental traces after treatment are given (blue-shaded) and compared to the theoretical spectra obtained from a simple electrostriction-photoelastic model.}
\end{figure}

\begin{figure*} \includegraphics[width=\textwidth]{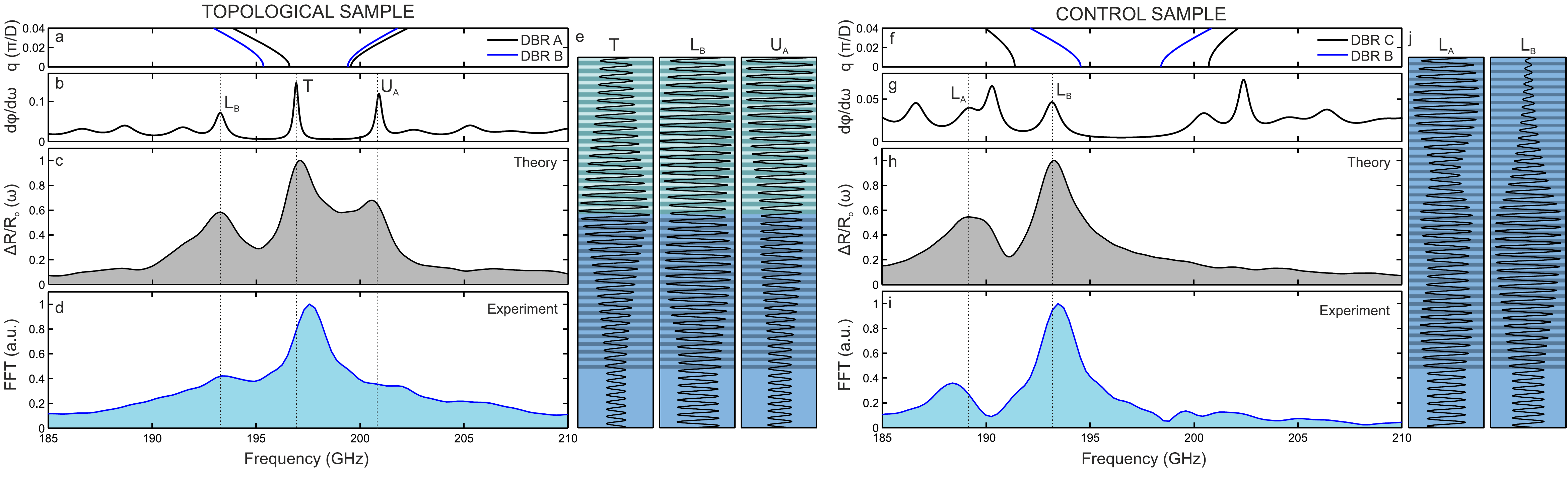}
    \caption{ \label{3} \textbf{Optical transient reflectivity spectra around 200 GHz}. (Left) (a) Band diagram of the underlying superlattices B and C. (b) Simulated derivative of the phase shift $\phi$ for a substrate-incident acoustic plane wave, allowing identification of supported resonances for a closed structure at the top layer. The three relevant modes are identified as $L_{B}$, $U_{A}$ and $T$ and are also present in the theoretical $\Delta R / R_o$ (c) and obtained experimental (d) spectra. (e) Mode profiles of the topological nanophononic cavity ($T$) as well as the detected lower (upper) band edge mode $L_{B}$ ($U_{A}$). (Right) Same for the control sample.}
\end{figure*}

\section{Phonon dynamics}

In order to access experimentally the phonon dynamics of the two samples presented in this work, we need to resolve physical processes at a picosecond time scale. For this purpose, we use a reflection-type pump-probe experiment at room temperature~\cite{thomsen}. The pump-probe technique relies on the usage of ultrafast laser pulses for both coherent phonon generation and detection. The experiment can be described as composed of two stages. First, a pump pulse is focused on the sample with enough power to impulsively photoinduce a stress $\sigma_{pump}(z,t)$ around the impact region, basically converting the optical energy into mechanical energy through various processes, generating a coherent acoustic phonon wave packet~\cite{ultrasonicsruellogusev}. Second, another, time-delayed pulse with significantly less power is used to probe the instantaneous reflectivity of the sample. The presence of acoustic excitations will modify the local optical properties, therefore modifying its reflectivity and allowing us to withdraw information of the phonons present in the nanostructure. By systematically sampling measurements at increasing delay times between pump and probe, we are able to monitor in time the transient optical reflectivity, therefore gaining access to the coherent dynamics of the generated phonons with temporal resolution given by the pulse length. Here, we used an asynchronous optical sampling (ASOPS) method, where two femtosecond Ti:sapphire lasers of repetition rate $f_{R}\sim$ 1 GHz are actively stabilized to have a repetition rate difference $\Delta f_{R}$ = 2 kHz. Such offset realizes the time delay between pump and probe pulse pairs, allowing a stroboscopic measurement transforming the time between two consecutive pump pulses ($\sim$ 1 ns) to 500 $\mu$s with increments of nearly 2 fs~\cite{bartelsASOPS}. The measurements were done at a fixed central wavelength of $\lambda$ = 800 nm for the pump beam (40 mW) and a varying central wavelength $\lambda$ = 760-840 nm for the probe (4 mW) beam, colinearly focused to a 2 $\mu m$ spot on the sample surface. Measurements at different magnifications and powers were done to rule out the presence of additional power density dependent temperature variations of the sample reflectivity. The Fourier transform of the differential reflectivity signals in the spectral region below 300 GHz is shown in Fig.~\ref{2} for both topological (top) and control (bottom) samples and compared to the expected differential reflectivity spectra extracted from a simple model based on electrostrictive forces for the generation and photoelasticity for the detection. The specific measurements shown Fig.~\ref{2} were performed at pump/probe central wavelengths of 800/760 nm for the topological sample and 800/800 nm for the control sample. In our simulations, we used the formalism presented in Refs.~\cite{generationmerlin} and~\cite{detectionmatsuda} assuming an impulsive generation mechanism. In order to reproduce the experimental conditions, the limited time window (1 ns) has been taken into account by convoluting the theoretical spectrum with a sinc function. The frequency cut-off induced by the finite size of the pulses ($\sim$50 fs) has also been taken into account, as is the full spectral width of the pump and probe pulses by doing a ponderated average of the generation and detection functions. It is clear from the time traces and the spectra in Fig.~\ref{2}(b) that specific coherent acoustic vibrations have been excited and that those can be read in the reconstructed differential reflectivity.\\

As shown above, a purely electrostrictive model for the phonon generation process and photoelasticity-based detection reproduces accurately the measured spectra for the samples and experimental conditions considered in this article. For GaAs/AlAs infinite superlattices in the transparency region, selection rules associated to such generation/detection mechanisms have been largely studied~\cite{mizoguchi,mfpascualwinter}. Generated coherent phonons correspond to forward scattering (FS) with $q=0$ phonons, i.e. zone-center acoustic excitations. Due to the well-defined symmetries of the Bloch modes at the band edges, an additional selection rule can be added: only modes with odd symmetry with respect to the bisecting plane ($A_{1}$ group symmetry) of the composing layers is accesible. The spectral response for detection is itself peaked at $q\sim2k_{eff}$, the effective wave-vector of the electromagnetic field. The spectral mismatch between the two processes is relaxed in finite realistic samples (DBRs) both by finite-size and absorption effects~\cite{mfpascualwinter}. When DBRs are used as building blocks for more complex multilayered structures, these simple selection rules are still extremely useful in understanding the observed spectra and allow for a qualitative understanding of the spectral components present in the transient reflectivities of Fig.~\ref{2}. Several acoustic modes are observed in Fig.~\ref{2}(b). The Brillouin peak~\cite{BrillouinPeak} of both superlattices and the substrate at $\sim$ 40 GHz, as well as two groups of peaks at 150 and 245 GHz, are precisely linked to the $q\sim2k_{eff}$ selection rule for the two different respective superlattices composing the samples. This is evidenced in Fig.\ref{2}(a), where the band structure of one of the superlattices (SL B) is depicted, with the horizontal line representing the $q=2k_{eff}$ condition at $\lambda$= 800 nm. The analysis of the region around the first zone-center minigaps of the underlying superlattices A, B and C (185-210 GHz) is depicted separately in Fig.~\ref{3}. The expected differential reflectivity spectrum $\Delta R / R_{o} (\omega)$ for the topological sample - Fig.~\ref{3}(c) - exhibits a modal structure involving three modes; owing to the inverted symmetry of the band edge modes of the underlying superlattices B and A and to the mentioned selection rules, the lower band edge mode of DBR B ($L_B$) and the upper band edge mode of DBR A ($U_A$) are present in the spectra. An additional central peak is associated to the topological mode ($T$) that arises precisely from this inversion of the symmetry. The derivative of the reflection phase from the substrate side depicted in Fig.~\ref{3}(b) and the calculated mode profiles $u(z)$ of Fig.~\ref{3}(e) clearly evidence the band edge and confined nature of the observed modes, respectively. The spectrum obtained after Fourier-transforming the differential reflectivity time trace obtained (Fig.~\ref{3}(d)) shows qualitative agreement with the theoretical spectra. The right side of Fig.~\ref{3} confirms that the simple selection rules for infinite non-absorbing superlattices are also conclusive for the control sample; the two peaks $L_C$ and $L_B$ corresponding to the lower zone-center band edge modes of DBRs C and B respectively.

\section{Conclusions}
We designed a topological nanoacoustic cavity with a resonance frequency of 200 GHz. The design is based on the inversion of the symmetries of the Brillouin zone edge modes of two concatenated superlattices. This band inversion relies on a small variation of the thicknesses of the constituting layers, due to the short wavelength of the considered phonons. We fabricated the sample and a control structure by MBE. We performed HRXRD characterization to confirm that the actual samples do not differ considerably from the nominal designs; i.e. the band structures associated to the actual thicknesses preserve acoustic band inversion. By means of pump-probe spectroscopy we characterized the photo-acoustic behavior of the two samples resulting in markedly different phononic spectra. The main features were remarkably well reproduced by transfer matrix simulations.

In a pump-probe measurement in a semiconductor superlattice we observe peaks that are related to $q=0$ and $q=2k_{eff}$ acoustic phonons. For symmetry reasons, only one of the two FS modes is accessible by the experiment. In the case of the control sample, two FS peaks are expected to appear on the same side of the common minigap. In the case of the topological cavity, the two FS peaks appear on opposite sides of the gap, validating the band inversion concept. Therefore, the use of pump-probe was essential to distinguish the three modes around 200 GHz. The central, most intense peak is direct proof for the existence of the topological mode. The possibility of identifying individual modes that are closely-spaced in frequency is of central importance for the study of dynamics in more complex topological acoustic structures where the interaction with the optical field can be engineered, as for example in topological resonators for light and acoustic phonons.

\section{Acknowledgements.}
ICN2 is supported by the Severo Ochoa program from Spanish MINECO (Grant No. SEV-2017-0706) and the project PHENTOM (Fis 2015-70862-P), as well as by the CERCA Programme / Generalitat de Catalunya, and by the European Commission in the form of the H2020 FET Open project PHENOMEN (GA. Nr. 713450). The authors acknowledge funding by the European Research Council Starting Grant No. 715939, Nanophennec; by the French RENATECH network, and through a public grant overseen by the ANR as part of the ``Investissements d\textsc{\char13}Avenir'' program (Labex NanoSaclay Grant No. ANR-10-LABX-0035). GA is supported by a BIST PhD fellowship and PDG by a Ramon y Cajal fellowship Nr. RyC-2015-18124. ME acknowledges funding from the German Research Foundation DFG (Forschungsstipendium ES 560/1-1).

\newpage

\end{document}